\documentstyle[prl,aps]{revtex}
\input epsf
\input psfig

   \newcommand{\be}{\begin{equation}}
   \newcommand{\ee}{\end{equation}}
   \newcommand{\bea}{\begin{eqnarray}}
   \newcommand{\eea}{\end{eqnarray}}
   \newcommand{\upar}{\uparrow}
   \newcommand{\dn}{\downarrow}

   \newcommand{\ar}{{\cal A}}
   \newcommand{\br}{{\cal B}}
   \newcommand{\om}{\omega_n}
   \newcommand{\bd}{\begin{displaymath}}
   \newcommand{\ed}{\end{displaymath}}

\begin{document}
\draft
\twocolumn[
\widetext
\title{Order from Disorder: Non Magnetic Impurities in the Spin-gap Phase of the Cuprates}
\author{Catherine P\'epin and Patrick A. Lee}
\address{Department of Physics, Massachussetts Institute of Technology, Cambridge, MA 02139}
\date{\today}

\maketitle \widetext
  \leftskip 54.8pt
  \rightskip 54.8pt
  \begin{abstract}
We solve the problem of $N$ non magnetic impurities in the staggered flux phase of the Heisenberg model which we assume to be a good mean-field approximation for the spin-gap phase of the cuprates. The density of states is evaluated exactly in the unitary scattering limit and is proportional to $1/\left (\omega \ln^2(|\omega|/D) \right )$, in analogy with the 1D case of doped spin-Peierls and two-leg ladder compounds. We argue that the system exhibits a quasi long-range order at $T=0$ with instantaneous spin-spin correlations decreasing as $n_i/ \ln^2\left (n_i R_{ij}\right )$ for large distances $R_{ij}$ and we predict enhanced low energy fluctuations in Neutron Scattering.
\par
  \end{abstract}
\vspace{0.2in}
]
\narrowtext

It is now established that the normal phase of the underdoped cuprates possesses a magnetic gap with the same anisotropy as the d-wave superconducting gap. The essential physics of the high-T$_C$ superconductors can be captured by focusing on the CuO$_2$ planes. A microscopic starting point of theoretical analysis is the so-called t-J model. In a slave-boson mean-field formulation, the system undergoes spin and charge separation: an electron in these highly correlated materials is a composite object made of a spin $1/2$ neutral fermion (spinon) and a spinless charged boson (holon). The gap in the magnetic excitations of the normal state can be viewed as singlet formation between pairs of neutral fermions in the absence of coherence between the holons. We assume that this mean-field picture captures essentially the initial idea of Anderson~\cite{anderson} of a Resonance Valence Bond ground state for the normal phase of the cuprates.
\par
Substitution of Cu ions in the conduction planes of high-T$_C$ cuprates by different non magnetic ions presents an important experimental tool for the study of the metallic state. Unusual effects has been revealed especially when the materials were doped with Zn. The valency of Zn is Zn$^{2+}$ (d$^{10}$) and compared with the Cu$^{2+}$ case, one electron is trapped by an additional positive charge of the nucleus, forming a singlet at the Zn site. In the spin-gap phase, it is experimentally found by NMR~\cite{xiao,alloul1,alloul2,mendels} experiments that a local magnetic moment of spin $1/2$ appears on Cu sites neighbouring the Zn impurity.
\par
In high-T$_C$ cuprates the persistance of AF fluctuations in the metallic state is probably one of the most striking feature. 
Inelastic Neutron Scattering have established previously the existence of an energy gap in the imaginary part of the dynamical susceptibility in the normal and superconducting phases of the pure compound~\cite{neutrons} (without Zn). Substitution of Zn in CuO$_2$ planes shows a striking transfer of spectral weight from high to low energies, partially filling the spin-gap~\cite{41mev,kakurai}. This is the signature of strong enhancement of AF correlations in the spin-gap phase.
\par

On the theoretical side, impurity-induced moments have been studied for a variety of quantum disordered spin systems. Especially in one dimension, doped spin ladders~\cite{nagaosasigrist,ladders1,laddersfukuyama,dagotto} such as SrCu$_2$O$_3$ and spin-Peierls~\cite{laddersfukuyama,peierls} systems such as CuGeO$_3$ have been investigated. These systems have in common with the underdoped cuprates the presence of a gap in the spin excitation spectrum. It has been shown theoretically as well as experimentally~\cite{expd1} that Zn doping is also the origin of a strong enhancement of AF correlations in the ground state.
The picture is that the impurity breaks up the singlet, leaving behind a $S=1/2$ local moment. Moments on the same and opposite sublattices interact with ferromagnetic and antiferromagnetic exchange respectively. There is no frustration in this system. The problem has been mapped onto the random $S=1/2$ chain with random sign interaction~\cite{nagaosasigrist} and is believed to exhibit long range or quasi-long range order at $T=0$. It has been suggested that similar physics may operate in two dimensional doped t-J model~\cite{laddersfukuyama}.
 
\par
Earlier work has treated the effect of a single non magnetic impurity in the $\pi$-flux state~\cite{autres1} and the pairwise interaction between them~\cite{autres2}. It was found that each impurity creates a bound state in the pseudogap at $\omega = 0$. Interaction between a pair of impurities leads to a level splitting between these states given by $\Delta(R) \sim1/\left ( R \ln (R) \right)$.

\par
Here we solve the problem of $N$ non magnetic impurities in the staggered flux phase of the Heisenberg model. We find that the overlap between the bound states leads to a broadening of the $\delta$-function peaks and the density of states is given exactly by
\bd
 \delta \rho \left ( \omega \right ) = n_i/ \left (| \omega | \ln^2 \left | \omega |/D \right ) \right ) \ ,
\ed ($n_i$ being the density of impurities). We also confirm that because the staggered flux phase introduces no frustration between the local moments, a quasi-long range staggered order exists at zero temperature. We present our results in terms of a $\pi$-flux phase even though our conclusions are general for any staggered-flux phase.
In our view the quasi long range
order found for the $\pi$-flux phase at half filling accounts for the general enhancement of antiferromagnetic correlations found by INS experiments in the underdoped compounds.

\par
In the absence of impurities, the Hubbard model
\be
 H= -t \sum_{\langle i,j \rangle, \sigma} (c^\dagger_i c_j + h. c.) + U \sum_i n_{i \upar} n_{i \dn}
\ee
for $U/t \gg 1$ at half-filling may be canonically transformed and projected into the nondoubly occupied subspace to become the $S=\frac{1}{2}$ AF Heisenberg model
\be
H^{AF}= -J \sum_{\langle i,j \rangle} \sum_{\sigma,\sigma^\prime} c^\dagger_{i \sigma} c_{j \sigma} c^\dagger_{j \sigma^\prime} c_{i \sigma^\prime}\; ,
\ee
where $J=4 t^2/U$. The mean-field approach assumes a nonzero expectation value for $\chi_{ij}= \langle c_{i \sigma}^\dagger c_{j \sigma} \rangle$, which is a link variable of a U(1) lattice gauge field coupled to the remaining fermions $\chi_{ij}=\exp [i \int_i^j {\bf A}. d {\bf l}]$. We parametrize the $\pi$-flux phase by taking $\chi_{ij}= \chi e^{i \theta_{ij}}$, where $\theta_{ij}= -\theta_{ji}$ and 
$\theta_{ij}=0$ if $ {\bf ij} \parallel \vec{x}$, and $\pm \frac{\pi}{2}$ if $ {\bf ij} \parallel \vec{y}$.

\par
Dividing the square lattice into two sublattices A and B, the mean-field hamiltonian can be diagonalized via the canonical transformation $
c^A_k= \psi_{k,0}+\psi_{k,1}$ and $
c^B_k= e^{i \varphi_k} (\psi_{k,0}-\psi_{k,1})$, 
with $ 
e^{i \varphi_k}= \left (\cos{k_x}+ i \cos{k_y}\right )/\sqrt{\cos^2{k_x}+ \cos^2{k_y}}
$, leading to $
H_{MF}= \sum_{k \in BZ,\alpha =0,1} E_k (-1)^\alpha \psi_{k,\alpha}^\dagger \psi_{k,\alpha}$,
 where the summation over $k$ extends over the full Brillouin zone and the constant terms have been omitted; $ \pm E_k = \pm \sqrt{\cos^2{k_x}+ \cos^2{k_y}} $ are the quasiparticle eigenvalues that will be linearized around the four nodes $(\pm \frac{\pi}{2},\pm \frac{\pi}{2})$.

\par
The impurities are treated as repulsive scalar potentials randomly distributed on the lattice
\be
H_{imp}= V_0 \sum_{i_0=1}^{N} c^\dagger_{i_0} c_{i_0} \ .
\ee
We assume that the unitary limit ($V_0 \gg 1$) is physical because the electrons on Cu sites are really trapped by Zn impurities in experimental realizations.

 We define the Green's function $G_{k k^\prime}^{\alpha \alpha^\prime} (i \om) =  -  \langle T_\tau  \psi_{k,\alpha}(\tau) \psi_{k^\prime, \alpha^\prime}^\dagger(0)  \rangle_{\omega_n}$, where the brakets $\langle \; \; \rangle_{\omega_n}$ denote the Fourier transform in Matsubara frequencies. Note that no impurity averaging has been made. The bare propagators for quasiparticles satisfy $ G_{k,\alpha}^0(i \omega_n)= \frac{1}{i \omega_n - (-1)^\alpha E_k}$.

The equation of motion for the Green's function can be closed algebraically: it involves $N$ unknown propagators (corresponding to the scattering by each impurities) that can be evaluated self consistently. This leads us to the following T-matrix equation
\bea
\label{eq1}
 &G_{k k^\prime}^{\alpha \alpha^\prime} &(i \om)  = G^0_{k,\alpha} \delta_{k,k^\prime} \delta_{\alpha,\alpha^\prime} \\
& & \; \; +  \sum_{i,j} {\cal G}^0_{k,\alpha}({\bf R}_i) \  T_{ij} (i \om) \  {\cal G}^0_{k^\prime, \alpha^\prime}(-{\bf R}_j) \nonumber \ ,
   \eea

where ${\cal G}_{k,\alpha}^0 ({\bf R}_i)= G^0_{k,\alpha} \exp \left (i {\bf k \cdot  R}_i \right )$ if ${\bf R}_i$ belongs to sublattice A and ${\cal G}_{k,\alpha}^0 ({\bf R}_i)=  G^0_{k,\alpha} \exp \left ( i {\bf k \cdot  R}_i -i \varphi_k \right )$ if ${\bf R}_i$ belongs to sublattice B. The T-matrix is given by ${\hat T} = - V_0 {\hat M}^{-1}$, with ${\hat M}$ the matrix that closes the equations of motion. ${\hat M}$ is a $N \times N$ matrix ($N$ is the number of impurities) each entry of which corresponds to the scattering from one impurity to another. It can be written by blocks as

\be
{\hat M} = \left [ \begin{array}{cc} {\hat A}_{N/2} & {\hat B}_{N/2} \\
      ^t{\hat B}_{N/2} & {\hat D}_{N/2} \end{array} \right ] 
\ee
where ${\hat A}$ and ${\hat D}$ describe the scattering by impurities in the same sublattice and ${\hat B}$ the scattering by impurities in different sublattices. We have noted in indices the rank of the matrices which is roughly $N/2$ but the solution works for any macroscopic number of impurities in each sublattice.
Dividing again in the same way each sublattice into two sub-sublattices A$_1$, A$_2$ and B$_1$, B$_2$, we define the coefficients
 $ {\hat A}_{ij}=  \delta_{ij} + V_0 \ \ar (i \om, {\bf R}_{ij}) $, if ${\bf R}_i$ and ${\bf R}_j$ belong to the same sub-sublattice and ${\hat A}_{ij}=0$ when ${\bf R}_i$ and ${\bf R}_j$ belong to different sub-sublattices.  ${\hat B}_{ij}= V_0 \ \br (i \om, {\bf R}_{ij}) $ and ${\hat D}_{ij}= {\hat A}_{ij}$. Two impurities in the same sublattice are related by the propagator $ \ar (i \om, {\bf R}_{ij}) = \frac{1}{{\cal N}} \sum_{k,n} G^0_{k,n} (i \om) \exp \left (i {\bf k \cdot  R}_{ij} \right ) $, with ${\cal N}$ the number of sites in the lattice.
When two impurities are in different sublattices, they are related by
$
\br (i \om, {\bf R}_{ij} ) = \frac{1}{{\cal N}} \sum_{k,n} G^0_{k,n} (i \om) (-1)^n \exp\left (-i \varphi_k + i {\bf k \cdot  R}_{ij} \right )$. Due to the symmetries of the $\pi$-flux phase, there is no interaction between two sub-sublattices that belong to the same sublattice.
\par

The coefficients $\ar (i \om, {\bf R}_{ij})$ and $ \br (i \om, {\bf R}_{ij})$ are given by
\bd
\ar(i \om, {\bf R}) \simeq \frac{4 i \om}{ \pi D^2} K_0\left(\frac{  R | \om|}{D} \right ) \  \Phi^A( R)  \ , \ed
where $\Phi^A( R ) = \frac{1}{2} \left (\cos(\pi R_x) +1 \right )$ and $K_0$ is the modified Bessel function of rank zero;
\bd 
\br( i \om, {\bf R} ) \simeq 4  i \exp\left[ i {\bf Q}_2 \cdot {\bf R} + i \varphi \right ]\Phi^B(R ) \ \frac{4 i \om}{ \pi D^2} K_1\left(\frac{  R | \om|}{D} \right ) \ ,\ed
with ${\bf Q}_2=( \frac{\pi}{2}, \frac{\pi}{2})$, $ \Phi^B(R) = \frac{1}{2} \left (1-\cos(\pi R_x) \right ) \exp(-2 i \varphi)$, $\varphi$ being the angle between ${\bf R}$ and $\vec{x}$~\cite{autres2}. 

 In the regime where $R_{ij} | \om | \ll D$,  $\ar_{ij}(i \om, {\bf R}) \sim 4/(\pi D^2) \ i \om \ln |R \om/D|$. This logarithmic structure is caracteristic of the system of Dirac fermions in 2 D. It is the origin of the breakdown of the usual perturbative expansions~\cite{alexei}: logarithmic divergences will indeed appear in all orders in the diagrammatic expansion, making it necessary to re-sum the whole series of diagrams in order to control any calculation. 

Here we take a non perturbative approach and directly invert the matrix ${\hat M}$. We notice that ${\hat A}_{N/2}$ approaches unity in the limit of low frequencies. We use
\be
\label{eq2}
\sum_k \br(i \om, R_{ik}) \br( i \om, R_{kj}) \sim \ln \left( | \om | R_{ij} /D \right) \ {\hat S} \ ,
\ee
where ${\hat S}$ is a $N/2 \times N/2$ matrix satisfying $S_{ij} =1$ if $R_{ij} < D/|\om|$ and $0$ elsewhere. The matrix product $^t{\hat B} {\hat B} \gg 1$. After taking the limit $V_0 \rightarrow \infty$, we invert the matrix ${\hat M}$ by blocks: 
\be
\label{eq3}
{\hat T}( i \om) = \left [ \begin{array}{cc}  0 & ^t{\hat B}^{-1} \\
                             {\hat B}^{-1}& 0 \end{array} \right ] \ .  
\ee
Impurities interact only if they are on different sublattices. We have supposed that ${\hat B}$ is a square matrix but the proof can easily be generalized for a non square one~\cite{cath}. 
\par
We proceed to evaluate the additional density of states $\delta \rho \left( \omega \right ) = - \frac{1}{\pi} \ Im \sum_{k, \alpha} \delta G_{kk}^{\alpha \alpha} \left( \omega + i \delta \right) $, where $\delta G$ is the part of the Green's function in eq.~(\ref{eq1}) due to impurities. It can be read from eq.~(\ref{eq1}) that $ \delta \rho \left ( \omega \right) =- \frac{1}{\pi}\  Im \  Tr \left ( {\hat T} \ \partial/\partial  \omega {\hat M} \right  ) $, so that $\delta \rho \left ( \omega \right) =- \frac{2}{\pi}\  Im \ \partial / \partial  \omega (  \ln  Det{\hat B} )$. The idea is now to evaluate this trace without exactly calculating the T-matrix~(\ref{eq2}). We know that $Det {\hat B}=1/2\  Det ( ^t{\hat B} {\hat B} )$ and we use the logarithmic form of $^t{\hat B} {\hat B}$~(\ref{eq2}) to get $Det {\hat B} \sim 1/2 \ f ( |\omega|) \ \ln^N \left ( | \omega | /D \right )$ where $f$ is a prefactor which doesn't depend on the number $N$ of impurities. The logarithmic structure of the problem that invalidates the usual perturbative expansions has been used here to factorize the leading divergence in the evaluation of $Det ( ^t{\hat B} {\hat B} )$. Since the prefactor $f(|\omega|)$ doesn't depend on $N$, its contribution to the density of states will be negligible in the thermodynamic limit. Hence in the thermodynamic limit and after analytic continuation, we get $\delta \rho \left ( \omega \right) = - \frac{N}{\pi} \ Im  \frac{\partial}{\partial  \omega} \left( \ln \ln \left| \frac{\omega}{D} \right | \right )$ and per unit of volume,

\be
\label{eq28}
\delta \rho( \omega) = \frac{1}{2 | \omega |}\  \frac{n_i}{ \ln^2(| \omega |/D ) + (\pi/2)^2} \ .
\ee

Note that $\int_{-\infty}^{\infty} \delta \rho(\omega) d \omega = n_i / 2$, there is thus as many states created at zero energy as impurities in the system. For one impurity alone~\cite{autres1,autres2}, a $\delta$-like bound state is created at $\omega=0$. If the impurities were totally uncorrelated, we would find $N$ $\delta$-functions at $\omega=0$. Here we see clearly the overlap of impurity states which leads to a broadening of the $\delta$ functions. It is interesting to compare this density of states with the one found for one dimensional spin-gap systems such as spin-Peierls~\cite{peierls} and two-leg ladders~\cite{ladders1,laddersfukuyama,nagaosasigrist} systems where ${\displaystyle \delta \rho(\omega) \sim 1/ \left(| \omega | \ln^3( |\omega|/D ) \right ) }$. It is also worth noticing that the very same form of the density of states as eq.~(\ref{eq28}) has been obtained in other apparently unrelated problem~\cite{brezin} for reasons which are not clear to us so far. 
\par

In our method, we in fact calculate exactly many thermodynamic quantities. For instance the uniform magnetic susceptibility is given by
\be
\chi(T)= \beta \int_0^{\infty} d\varepsilon \delta \rho(\varepsilon) \frac{1}{2 \cosh^2(\frac{\beta \varepsilon}{2})} \ .
\ee

From the asymptotic behavior of $\delta \rho(\varepsilon)$ at small $\varepsilon$ we find that at low temperatures ${\displaystyle \chi(T) \sim n_i / \left ( 2 T \ln (1/T) \right ) }$, which can be interpreted as a Curie-like behavior with a vanishing Curie constant ${\displaystyle C(T)=n_i/ \left ( 2 \ln(1/T) \right ) }$.

The free-energy 
\bd
\Delta F(T)= -2 T \int_0^\infty d\varepsilon \delta \rho(\varepsilon) \ln ( 1+ e^{- \beta \varepsilon} )
\ed leads to a specific heat ${\displaystyle C_V(T) \sim 1/(\ln^2(1/T))}$ and a low temperature entropy ${\displaystyle S(T) \sim 1/\ln(1/T) }$. However the entropy vanishes very slowly with temperature. 

\par
We address now the important question of long range order in the system by computing the instantaneous correlations $\langle S_i^+ S_j^- \rangle= \lim_{\tau \rightarrow 0^+} \left \langle S_i^+(\tau) S_j^-(0) \right \rangle$. It is easy to show that $\langle S_i^+ S_j^- \rangle = \lim_{\tau \rightarrow 0^+} \langle G_{ij}^{\upar}(\tau) G_{ji}^{\dn}(-\tau) \rangle$, where $ G_{ij}(\tau)$ is the Fourier transform in real space and complex time of the exact Green's function of eq.~(\ref{eq1}).  We verify that this two-spin correlation is staggered. If $i$ and $j$ belong to the same sublattice, $\delta G_{ij}(i \om) = \sum_{kl} \ar(i \om, {\bf R}_{ik})\  T_{kl}\  \br( i \om, {\bf R}_{lj})$ which owing to~(\ref{eq3}) simply renormalizes the bare Green's function: $\delta G_{ij}(i \om) = -n_i \ar(i \om, R_{ij})$. When $i$ and $j$ belong to different sublattices, there is one non trivial quantity $ \delta G_{ij}(i \om) = \sum_{kl} \ar(i \om, {\bf R}_{ik})\  T_{kl}\  \ar( i \om, {\bf R}_{lj})$. A simplification is obtained after noticing that ${\hat A} = {\hat B}^{-1} \ (\partial {\hat B} / \partial i \om) $, which leads to 
\be
\label{eq4}
 \delta G_{ij}(i \om) =  \left( \partial {\hat B}  / \partial i \om  \right ) \left( \partial {\hat B} ^{-2} / \partial i \om  \right) \ .
\ee
 Writing $ \partial{\hat B}^{2} / \partial i \om  = 2 {\hat B} \ ( \partial{\hat B} / \partial i \om )$ and evaluating this product with the use of eq.~(\ref{eq2}), we show that $ \partial{\hat B}^{2} / \partial i \om  \sim \left( \partial \ln (| \om | /D) / \partial i \om  \right ) \  {\hat S}$. Here again we have factorized the logarithmic divergence in ${\hat B}^2$ and shown that the derivative with respect to $i \om$ acts mainly on this factorized logarithm. Consequently $ \partial{\hat B}^{-2} / \partial i \om  \sim \left ( \partial ( 1/\ln|\om|/D) / \partial i \om   \right ) \  {\hat S}^{-1}$. We now make an approximation for the inverse of ${\hat S}$. We take ${\hat S}^{-1}_{ij} = (|\om|/D) \ U_{ij}$. For all $j$ inside a circle of radius $D/|\om|$ around the point $i$, $U_{ij}$ is a random configuration of $\pm 1$ so that $\sum_j U_{ij} \sim  D/|\om|$. In addition all the points $j$ situated in the external boundary of this circle have $U_{ij}= -1/\pi$. Elsewhere $U_{ij}=0$. The main difficulty in the inversion of ${\hat S}$ is that two circles of radius $D/|\om|$ centered around two points $i$ and $j$ very close to each other will overlap, leading to the same number of non zero coefficients in the lines $i$ and $j$ of ${\hat S}$. In order to differentiate the sums $\sum_k S_{ik} S_{ki}^{-1}$ and $\sum_k S_{ik} S_{kj}^{-1}$ we have thus used the external boundary of the circle to compensate its volume in ${\hat S}^{-1}$. Using this inverse we perform the evaluation of eq.~(\ref{eq4}). In order to take into account the vertex corrections due to scattering processes, we perform the average over disorder after having evaluated the product of the two Green's functions. We find that at $T=0$, 
\be
\label{eq20}
\langle S_i^+ S_j^- \rangle \sim \frac{n_i}{\left ( \ln^2\left ( n_i R_{ij} \right ) + (\pi/2)^2 \right) }\ ,
\ee
for impurities in different sublattices.
 $N$ non magnetic impurities randomly distributed in a $\pi$-flux phase stabilize a quasi long range staggered order. This result has to be compared with the pure $\pi$-flux instantaneous correlations which decay algebraically in $1/R^4$~\cite{laughlin} at zero temperature. Physically we understand what happens by introducing impurities one after another. For one impurity, a local moment with $S=1/2$ is created in the vicinity~\cite{autres1}. Under on-site repulsion, two impurities interact via an effective Heisenberg exchange $H=J(R) {\bf S}_1 . {\bf S}_2$, with $J(R) \sim 1/(R \ln(R))$ if they are located on different sublattices. However two impurities located on the same sublattice don't interact~\cite{autres2}. In our model, eventhough the on-site repulsion is treated at the mean-field level, the one particle wave-function remains commensurate with sublattices. This supports the intuition that no frustration will be introduced in the model when a macroscopic number of impurities interact with each other. This effect is the origin of the quasi long range order that we find.

A similar picture is obtained for the ladder or spin-Peierls cases except that in these cases the instantaneous spin correlation decreases exponentially with density, reflecting the fact that interactions between local moments fall exponentially with the distance. Here the instantaneous correlations on different sublattices are proportional to the density of impurities because a pair of impurities interact via an effective exchange $J(R) \sim 1/(R \ln(R))$. We suggest that the logarithmic decay in~(\ref{eq20}) is due to quantum fluctuations at zero temperature even though the precise power in the logarithm may depend on our approximation.
\par
What would be observed if a Neutron Scattering experiment would be performed on this system?  We have computed here the structure factor ${\bar S} ({\bf Q}, \Omega) = 1/{\cal N} \sum_q \chi^{\prime \prime}({\bf Q}, \Omega)$ integrated around $(\pi, \pi)$. For two points $i$ and $j$ in different sublattices, $\chi^{\prime \prime}_{ij}({\bf Q}, \Omega)$ is expressed as
$
\chi^{\prime \prime}_{ij}({\bf Q}, \Omega+ i \delta)= 1/ (\pi \beta) \int_0^\Omega d\omega \ Im G_{ij}( \omega -\Omega +i \delta ) \ Im G_{ji}( \omega  +i \delta ) 
$.
Using the same tricks as before for the instantaneous correlation function, we show that at low frequencies
\be
\label{eq13}
{\bar S}({\bf Q}, \Omega) \sim \frac{n_i}{\Omega \ln^4 \left | \Omega / D \right | } \ .
\ee
Even though we don't have true long range order in the system (which would have led to ${\bar S}({\bf Q}, \Omega) = \delta (\Omega)$ ), Neutron Scattering experiments would see a divergence of the integrated staggered structure factor at low frequencies, signaling a strong enhancement of antiferromagnetic correlations.

The situation will change for experimental systems with finite oxygen doping. If we take as a theoretical starting point the d-RVB ground state, the effect of oxygen doping will be to move the nodes away from the four points $(\pm \pi/2,\pm \pi/2)$ in the canonical spectrum. Excitations from node to node will then slightly differ from the commensurated vectors $(\pi,\pi)$. We believe that instead of quasi-long range AF order, this mechanism may give rise to an incommensurate quasi-long range order. Furthermore, it is likely that the finite frequency response integrated over $q$-space will not be that different from that given by eq.(\ref{eq13}), which may explain the enhancement of AF correlations observed in Zn doped YBaCuO by neutron scattering~\cite{41mev,kakurai}. 

It is a pleasure to acknowledge useful discussions with J. Brinckmann, C. Mudry, Yu Lu, N. Nagaosa, A. Furusaki, T.K. Ng and C. Castellani.  One of us (CP) is deeply indepted to E. Kowalski for great help with linear algebra and complex analysis.
 This work is supported by NSF Grant No. DMR-9523361 and by (CP) a Bourse Lavoisier.





\begin{references}

\bibitem{anderson} P. W. Anderson, Science {\bf 235}, 1196 (1987). 
\bibitem{neutrons} J. Rossat-Mignot {\it et al.}, Physica C {\bf 185-189}, 86 (1991) or H. F. Fong {\it et al.}, Phys. Rev. Lett. {\bf 75}, 316 (1995) or P. Bourges {\it et al.}, Phys. Rev. B {\bf 53}, 876 (1996).
\bibitem{xiao} G. Xiao {\it et al.}, Phys. Rev. B {\bf 35}, 8782 (1987).
\bibitem{alloul1} H. Alloul {\it et al.}, Phys. Rev. Lett. {\bf 67}, 3140 (1991).
\bibitem{alloul2} A. V. Mahajan {\it et al.}, Phys. Rev. Lett. {\bf 72}3100 (1994).
\bibitem{mendels} P. Mendels {\it et al.}, Phys. Rev. B {\bf 49}, 10035 (1994).
\bibitem{41mev} Y. Sidis {\it et al.}, Phys. Rev. B {\bf 53}, 6811 (1996).
\bibitem{kakurai} K. Kakurai {\it et al.}, Phys. Rev. B {\bf 48}, 3485 (1993).
\bibitem{nagaosasigrist}N. Nagaosa {\it et al.}, JPSJ {\bf 65}, 3724 (1996); M. Sigrist and A. Furusaki, JPSJ {\bf 65} (1996).
\bibitem{ladders1} A. O. Gogolin {\it et al.} cond-mat/ 9707341.
\bibitem{laddersfukuyama} H. Fukuyama {\it et al.}, JPSJ {\bf 65}, 2377 (1996);JPSJ {\bf 65}, 1182 (1996).
\bibitem{dagotto} G. B. Martins {\it et al.} Phys. Rev. Lett. {\bf 78}, 3563 (1997).
\bibitem{peierls}M. Fabrizio and R. M\'elin, Phys. Rev. Lett. {\bf 78}, 3382 (1997).
\bibitem{expd1} J. P. Renard {\it et al.}, Europhys. Lett. {\bf 30}, 475 (1995); L-P. Regnault {\it et al.}, Europhys. Lett. {\bf 32}, 579 (1995).
\bibitem{autres1} N.Nagaosa and T.K Ng, Phys. Rev. B {\bf 51}, 15588 (1995); N.Nagaosa and P. A. Lee, Phys. Rev. Lett. {\bf 79}, 3755 (1997).
\bibitem{autres2} S. Krivenko and G. Khaliullin, JETP Lett. {\bf 62}, 723 (1995); G. Khaliullin {\it et al.} Phys. Rev. B  {\bf 56}, 11882 (1997).
\bibitem{alexei} A.A Nersesyan, A. M. Tsvelik and F. Wenger, Nucl. Phys. B {\bf 438}, 561 (1995).
\bibitem{cath} We transform ${\hat B}$ into a square matrix by adding some columns or lines of $\varepsilon$  with $\varepsilon \rightarrow 0$ to ${\hat B}$ and the same number of 1 in the diagonal of ${\hat M}$.
\bibitem{brezin}  E. Br\'ezin, D.J. Gross and C. Itzykson, Nucl. Phys. B {\bf 235} (1984), 24-44.
\bibitem{laughlin} R. B. Laughlin and Z. Zou, Phys. Rev. B {\bf 41}, 664 (1989).

\end{references}
\end{document}